\def\lea{\mathrel{<\kern-1.0em\lower0.9ex\hbox{$\sim$}}}
\def\gea{\mathrel{>\kern-1.0em\lower0.9ex\hbox{$\sim$}}}

 \documentclass[preprint]{aastex}

\slugcomment{To appear in The Astronomical Journal}

\shorttitle{Red Clump Distances to LMC Clusters}
\shortauthors{Sarajedini et al.}

\begin{document}


\title{K-Band Red Clump Distances to the LMC \\ Clusters Hodge 4 and NGC 1651}


\author{Ata Sarajedini, Aaron J. Grocholski\altaffilmark{1}, Joanna Levine\altaffilmark{1},  
and Elizabeth Lada}
\affil{Department of Astronomy, University of Florida, Gainesville, FL 32611}
\email{ata@astro.ufl.edu, aaron@astro.ufl.edu,levine@astro.ufl.edu,lada@astro.ufl.edu}


\altaffiltext{1}{Visiting Astronomer, Cerro Tololo Inter-American Observatory.
CTIO is operated by AURA, Inc.\ under contract to the National Science
Foundation.}


\begin{abstract}
We present near-infrared color-magnitude diagrams for the LMC star clusters Hodge 4 and 
NGC 1651 revealing the helium burning red clump stars for the first time in the JK passbands. 
From these diagrams, which extend to $K\sim19$, and existing optical CMDs that reveal
the clusters' main sequence turnoffs, we determine the following cluster parameters.
For Hodge 4, we estimate a metallicity of $[Fe/H] = -0.17 \pm 0.04$, an age of
$1.7 \pm 0.3$ Gyr, and a red clump absolute magnitude of $M_K(RC) = -1.64 \pm 0.17$.
Along with an adopted reddening of $E(J-K) = 0.03 \pm 0.01$ based on the Burstein \& Heiles
and Schlegel et al. reddening maps, we find a distance of $(m-M)_0 = 18.52 \pm 0.17$ for Hodge 4. 
In the case of NGC 1651, we derive $[Fe/H] = -0.07 \pm 0.10$, an age of
$1.8 \pm 0.3$ Gyr , $M_K(RC) = -1.56 \pm 0.12$, $E(J-K) = 0.06 \pm 0.01$, and
$(m-M)_0 =18.55 \pm 0.12$, all measured in the same manner as for Hodge 4.
Based on these two clusters, we calculate a mean LMC distance of 
$(m-M)_0 = 18.54 \pm 0.10$.
\end{abstract}




\section{Introduction}

The intrinsic colors and luminosities of stars are a powerful probe of their properties, 
such as metal abundance and age. For sources not embedded in significant amounts of dust, 
the traditional choice has been the use of optical colors (e.g. $B-V$, $V-I$) and 
magnitudes for this purpose. Because of this, there exists a substantial body of 
foundational work that attempts to link stellar photometry in the optical regime to 
fundamental stellar parameters.  It is for this reason that in many applications, observations 
in the optical are the preferred method by which to glean information about stars in 
unobscured regions.

There is at least one notable exception to this assertion, however. The use of red clump (RC)
stars as distance indicators, which has utilized observations in the 
V-band (Sarajedini, Lee, \& Lee 1995), R-band (Seidel, Da Costa, \& Demarque 1987),
and I-band (Paczy\'{n}ski \& Stanek 1998), has lately moved into the infrared wavelengths 
(Alves 2000; Grocholski \& Sarajedini 2001, hereafter GS01; Alves et al. 2002); 
this is due to the realization that the K-band absolute
magnitude of the red clump [$M_K(RC)$]  is less susceptible to uncertainties in the age 
and abundance of the stellar population as compared with its optical cousins. 

The work of GS01 used $JK_s$ photometry for Galactic open clusters from the Two Micron 
All Sky Survey (2MASS) to establish a calibration between cluster age, metallicity, and the
K-band absolute magnitude of the red clump. The empirical calibration was closely mirrored
by the theoretical predictions based on the models of Girardi et al. (2000). Given the usefulness
of such a robust distance calibration, GS01 sought to apply it to a cosmologically significant
problem such as the distance of the Large Magellanic Cloud (LMC). However, it quickly became
apparent that there is a dearth of near-IR photometry for helium burning red clump stars in 
LMC clusters. All of the extant near-IR studies of intermediate age clusters in the LMC 
(i.e. older than $\sim$1 Gyr)  focus on stars with $K\lea16$ (e.g. Aaronson \& Mould 1985,
and references therein; 
Ferraro et al. 1995), whereas the red clump is expected to be at $K\sim17$. 

The present work represents our initial efforts to remedy this situation. We have obtained JK 
imaging for the LMC clusters Hodge 4 and NGC 1651. Sections 2 and 3 describe the observations
and data reduction procedures used to construct the near-IR color-magnitude diagrams (CMDs)
discussed in Sec. 4. The cluster metallicities, ages, reddenings, and distances are discussed in Section 5.  
Section 6 presents a comparison between the observations and the theoretical models of the red 
giant branch. Section 7 contains a summary of our results.

\section{Observations}

J-band (1.22$\mu$m) and K$_{s}$-band (2.16$\mu$m) images of Hodge 4 
($\alpha_{2000}$ = $5^h$32'26",  $\delta_{2000}$ = --$64^o$44'11") 
and NGC 1651 ($\alpha_{2000}$ = $4^h$37'32",  $\delta_{2000}$ = --$70^o$35'06") 
 were obtained over 3 nights in December 2001 using the Cerro
Tololo Inter-American Observatory (CTIO) 4m telescope with the Ohio State
InfraRed Imager/Spectrometer (OSIRIS), in conjunction with the tip-tilt
image stabilization system.  OSIRIS uses a 1024$\times$1024 pixel mercury
cadmium telluride (HgCdTe) HAWAII focal plane array.  All data were
obtained with the f/14 secondary and the OSIRIS f/7 camera. This optical
configuration yielded a field of view of 93\arcsec$\times$93\arcsec and a
plate scale of 0\farcs161/pixel. Wavefront reference was provided by one
of two optically visible guide stars within 4' of each of the targeted
science fields.

We observed two fields in each cluster at J and K$_{s}$: one centered on
the target and one offset by 60\arcsec to the east or west.  The total area
covered for each cluster was $\sim$3.75 square arcminutes.  All data were
taken using a 9-point dither pattern. For the J-band data, the integration
time at each point was 11 seconds with 2 coadds, resulting in a total of
198 seconds on source. For data taken at K$_{s}$, the integration time at
each point was 10 seconds with 3 coadds, yielding a total of 270 seconds
on source.  Typical seeing at the telescope was $\sim$0\farcs6.

Figure 1 shows our observed fields in Hodge 4 (1a) and NGC 1651 (1b). 
Each image is approximately 150" x 90" in size. These clusters are
located at substantial distances from the center of the LMC -- 4.9$^o$ in the
case of Hodge 4 and 4.2$^o$ for NGC 1651. Figure 1 suggests that
both clusters are situated in relatively sparse regions with little background
contamination from LMC field stars.

\section{Reductions}

Images were processed with a reduction pipeline composed of a number of IRAF 
 \footnote{IRAF is distributed by the National Optical Astronomy Observatory,
which is operated by the Association of Universities for Research in Astronomy,
Inc., under cooperative agreement with the National Science Foundation.}
scripts, which performed the following tasks. The nine dithered source images were 
dark subtracted
and median combined, so as to eliminate any objects on each frame,
leaving only an exposure of the sky.  A flat field image was created by
normalizing this sky frame.  Object frames were then sky subtracted and
flat fielded using these images.  After processing, spatial offsets were
calculated and the images were shifted and average combined, creating a
single image for each observed field with signal-to-noise ratio
equivalent to a single 198s exposure in J and a single 270s exposure in
$K_s$.

\subsection{Standard Stars}

The combined standard star  images were analyzed with the IRAF/QPHOT software
package, which was used to measure instrumental magnitudes using a 15 pixel radius 
aperture.  We combined standard star measurements from all three 
nights by offsetting all of the measured instrumental magnitudes to the same 
zeropoint using stars in common between the three nights.  
The combined dataset of 7 standard star measurements
was then fit using a least-squares algorithm
to determine the extinction coefficients and zeropoints.  This procedure yielded the
following equations:
\begin{eqnarray}
j-J = (2.03 \pm 0.03) + (0.06 \pm 0.02) \times X_J
\end{eqnarray}
\begin{eqnarray}
k_s - K_s = (2.23 \pm 0.03) + (0.09 \pm 0.02) \times X_K
\end{eqnarray}
\noindent as shown in Fig. 2. In this formulation, lower case letters represent 
exposure-time-corrected instrumental magnitudes
and X is the airmass. The root-mean-square deviation of the points from the fitted relations was 
0.014 mag in $J$ and 0.013 mag in $K_s$. 

\subsection{Program Clusters}

The combined $JK_s$ images of Hodge 4 and NGC 1651 were measured using
the aperture photometry routines in DAOPHOT (Stetson 1994). Magnitudes
were determined in a 3 pixel-radius aperture for all detected stellar profiles. These
were then corrected to total instrumental values via the application of an aperture
correction derived from the brightest uncrowded stars in each field. We also measured 
the brightest stars in each frame 
using the IRAF/QPHOT task, which was used for the standard star reductions.
The DAOPHOT and QPHOT total aperture magnitudes
agreed to within 0.01 mag. Furthermore, photometry for stars located in the overlap
region between the two fields observed in each cluster differed by less than 0.02 mag
as compared with the mean photometric zeropoint.

The next step in the photometric reduction involved combining the instrumental 
magnitudes for stars in common between the two fields in each cluster; this was done
with the DAOMASTER (Stetson 1994) software. The resulting photometry was corrected for
exposure time and transformed to the LCO standard system of Persson et al. (1998)
using Eqs 1 and 2 above. The final step involved transforming our photometric system 
to that of Bessell \& Brett (1988) via Eqs. 26 and 27
of  Carpenter (2001) in conjunction with Eqs. 1 and 2 of GS01. This was
done in order to facilitate the use of the red clump distance calibration established by GS01.

\section{Color-Magnitude Diagrams}

Figure 3 shows the ($K,J-K$) color-magnitude diagrams (CMDs) of Hodge 4 and 
NGC 1651. These contain all stars that were photometered in the two fields.
The same basic
features are apparent in each CMD. There is a prominent red clump located at
K$\sim$17 with a well-populated red giant branch extending $\sim$4 magnitudes
brightward of the RC. The few stars located just above the RC are likely to be post
horizontal branch (i.e. red clump) stars. In the NGC 1651 CMD, there are two conspicuous
bright stars that extend above the first ascent red giant branch tip. In Sec. 6, we assert that
these are likely to be asymptotic giant branch stars. 

\section{Cluster Properties}

\subsection{Metallicities}

The most straightforward technique by which to determine cluster metallicities using
CMDs is via
the slope of the RGB. The slope becomes shallower as a clusters' metal abundance 
increases (Kuchinski et al. 1995; Ferraro et al. 2000). 
In order to measure this slope, we apply
an iterative 2-$\sigma$ rejection linear least squares fitting routine to stars brightward and 
redward of the red clump (Sarajedini \& Norris 1994). The resultant fits and the fitted
points are shown in the two left panels of Fig. 4.  We find RGB slopes of $-0.095 \pm 0.003$ 
for Hodge 4 and $-0.102 \pm 0.007$ for NGC 1651. The quoted uncerainties are the formal
least squares fitting errors.

The work of Tiede, Martini, \& Frogel (1997) has established a relation between RGB
slope and metal abundance for Galactic open clusters, which are similar in age to
the LMC clusters considered herein. We have recalibrated the zeropoint of their relation
using the metallicities published by Twarog et al. (1997), which is the abundance scale 
used by GS01. We find

\begin{eqnarray}
[Fe/H] = (-1.519 \pm 0.073) + (-14.243 \pm 1.963) \times RGB ~ Slope,
\end{eqnarray}

\noindent where the slope is that derived by Tiede et al. and the zeropoint is the one appropriate for
the Twarog et al. metal abundance scale.
The RGB slopes given above then yield metal abundances of $[Fe/H] = -0.17 \pm 0.04$
for Hodge 4 and $[Fe/H] = -0.07 \pm 0.10$ for NGC 1651. The uncertainties in these quantities
represent the error in the slope propagated through the calculation.

These near-IR RGB slope-based abundances are about the same or slightly higher than 
previous estimates. For example, Olszewski et al. (1991) present Calcium triplet spectroscopy of giant
stars in a number of LMC clusters. They find $[Fe/H] = -0.15$ for Hodge 4 based on observations
of one star and a somewhat uncertain value of $[Fe/H] = -0.37$ for NGC 1651 based on
four stars. Previous photometric studies of these clusters in the optical
tend to find significantly lower abundances. For 
example, Mateo \& Hodge (1986) apply various techniques to their UBV CCD photometry to 
arrive at a mean metallicity of $[Fe/H] = -0.7 \pm 0.3$ for Hodge 4. In the case of NGC 1651,
Mould, Da Costa, \& Crawford (1986) use the dereddened color of the RGB at the level of the
RC to derive $[M/H] = -0.5 \pm 0.3$, while Mould et al. (1997) apply a maximum likelihood 
isochrone fit to HST photometry (see next Section) to arrive at  $[Fe/H] = -0.4$.  

\subsection{Ages}

With the metallicities in hand from the previous section, we can move on to estimating
ages for the clusters in the present investigation. The most reliable technique for this
purpose is to examine the luminosity of the main sequence turnoff and compare it to the
predictions of theoretical isochrones. We are well-placed to perform such a comparison
because both Hodge 4 and NGC 1651 possess CCD photometry from the Wide Field
Planetary Camera 2 (WFPC2) on the Hubble Space Telescope (HST). 

In the case of Hodge 4, Sarajedini (1998) has published a $BV$ CMD that reaches
V$\sim$23. These data were taken in 1994 February before WFPC2 was cooled to its
nominal operating temperature, which means that the resulting photometric zeropoints 
are especially uncertain. In light of this fact, we proceed as follows. Using the theoretical
isochrones from Girardi et al. (2000) for Z=0.008 ($[M/H] = -0.4$) and Z=0.019
($[M/H] = 0.0$), we shift them vertically to match the red clump magnitude of Hodge 4
and horizontally to match its unevolved main sequence. The resulting fits are shown in
Fig. 5a. The upper panel suggests an age of 1.6 Gyr while the lower panel indicates 
1.8 Gyr. When we interpolate to the metallicity we determined earlier, we find that Hodge 4 has
an age of $1.7 \pm 0.3$ Gyr. This agrees reasonably well with the previous determinations
from Mateo \& Hodge (1986) who found an age of $2.0 \pm 0.5$ Gyr and Sarajedini
(1998) who also estimated an age of 2.0 Gyr. We note that both of these studies determined
the age of Hodge 4 based on an adopted metal abundance of $[Fe/H] = -0.7$ which is 
significantly lower than our value. Using a higher metal abundance would decrease
their quoted ages.

For NGC 1651, we utilize the HST WFPC2 data published by Mould et al. (1997). As in
the case of Hodge 4, we use the Girardi et al. (2000) isochrones for Z=0.008 and Z=0.019;
we shift the isochrones vertically to match the magnitude of the red clump and
horizontally to match the unevolved main sequence. These comparisons are shown in the
two panels of Fig. 5b. The upper panel suggests an age of 1.6 Gyr while the lower panel indicates 
1.8 Gyr. Given the calculated metal abundance of 
$[Fe/H] = -0.07 \pm 0.10$ for NGC 1651, the isochrones suggest an age of $1.8\pm0.3$
Gyr. This compares favorably with the two previous CMD studies of this cluster. Mould
et al. (1986) find an age of $2.0\pm0.8$ Gyr based on $BR$ 
photometry; Mould et al. (1997) find an age of $1.6\pm0.4$ Gyr based on a
maximum likelihood isochrone fit to their HST photometry, which we have used to perform our isochrone
comparisons.

\subsection{Reddenings}

Both Hodge 4 and NGC 1651 are located at substantial distances from the LMC bar. As a result,
we do not expect them to suffer from significant internal extinction due to the LMC itself. The 
Burstein \& Heiles (1982) maps
yield $E(B-V) = 0.035$ and $E(B-V) = 0.100$ for Hodge 4 and NGC 1651, respectively, while
the dust maps of Schlegel et al. (1998) give $E(B-V) = 0.059$ and $E(B-V) = 0.140$. These
values are sufficiently close to each other that we decided to average them to arrive at our
adopted reddening values of $E(B-V) = 0.05 \pm 0.01$ and $E(B-V) = 0.12 \pm 0.02$ for
these clusters. Using the relations quoted by Schlegel et al. (1998), 
$A_V = 3.1 E(B-V)$, $A_K = 0.11A_V$, $A_J = 0.28A_V$, these reddenings translate to 
$E(J-K) = 0.03 \pm 0.01$ and $E(J-K) = 0.06 \pm 0.01$ for Hodge 4 and NGC 1651.

We can also estimate the cluster reddenings using the intrinsic color of the red clump  
parameterized in terms of the metallicity and age as shown in Fig. 8 of GS01. Doing this, we find 
$(J-K)_0 = 0.600 \pm 0.006$ for Hodge 4 and $(J-K)_0 = 0.612 \pm 0.019$ for NGC 1651.
Combining these with the median observed colors of the red clumps, $(J-K) = 0.627 \pm 0.006$ 
and $(J-K)_0 = 0.684 \pm 0.006$, respectively, we calculate reddenings of 
$E(J-K) = 0.03 \pm 0.01$ and $E(J-K) = 0.07 \pm 0.02$ for Hodge 4 and NGC 1651.
Both of these values are in good agreement with the predictions of the reddening maps
discussed above.

\subsection{Distances}

With the ages, metallicities, and reddenings in hand, we can proceed to estimate the distance to 
each cluster. The K-band apparent magnitude of the red clump [$K(RC)$] is measured using 
the median of all stars in the boxes shown in Fig. 4,
with an error estimated as the standard error of the mean. This is the same technique employing
the same size boxes as GS01. The right panels of Fig. 4 display the K-band luminosity function for
each cluster clearly illustrating the appearance and location of the RC. We note that, to remain
consistent with GS01, whose distance calibration we utilize below, we have not adopted
the Gaussian fitting method to estimate $K(RC)$ as others have advocated 
(e.g. Paczy\'{n}ski \& Stanek 1998). 
We also require a value for the absolute magnitude of the red clump in each cluster 
($M_K(RC)$).
This is provided by the calibration of GS01 given the derived values of the age and metal
abundance. The results of this procedure are given in Table 1. The quoted error in $M_K(RC)$ 
contains contributions from the errors in metal abundance and age both added in quadrature.
In particular, the contributions are 0.047 and 0.066 mag for Hodge 4 and NGC 1651, respectively,
due to the metal abundance errors and 0.166 and 0.101 mag as a result of the age errors.

We note that the errors on $M_K(RC)$ due to age uncertainties are greater than they would 
otherwise be because the ages of Hodge 4 and 
NGC 1651 place them at an inflection point in the relation that governs the behavior of
$M_K(RC)$ with age (Fig. 6 of GS01). All else being equal, for ages older than $\sim$2 Gyr, 
the typical error on $M_K(RC)$ is less than 0.1 mag. In any case, the distances of Hodge 4
and NGC 1651 agree reasonably well with each other and lead to a weighted mean value of 
$(m-M)_0 = 18.54 \pm 0.10$ corresponding to $51\pm3$ kpc. To within the errors, this agrees 
with the mean
value of $(m-M)_0 = 18.50 \pm 0.04$ quoted by van den Bergh (1999) based on a variety
of distance indicators and the distance determined by Alves et al. (2002) of 
$(m-M)_0 = 18.49 \pm 0.04$, which is based on the K-band magnitude of the red clump among
LMC field stars.  We note in passing that adopting cluster metal abundances that are 0.3 dex lower
than those given in Table 1 would increase our mean LMC distance by $\sim$0.05 mag.

\section{Near-IR Isochrones}

The work of GS01 showed that the K-band red clump absolute magnitudes predicted by the 
theoretical models of
Girardi et al. (2000) agree well with the observed values. With the well-populated RGBs of
Hodge 4 and NGC 1651, we can test how well the JK theoretical {\it RGBs} fit the observed ones. 
Like the study of GS01, we utilize the Girardi et al. (2000) models, which are already on the 
Bessel \& Brett (1988) photometric system. Figure 6 shows comparisons between our cluster
photometry and the Z=0.008 ($[M/H] = -0.4$) and Z=0.019 ($[M/H] = 0.0$) isochrones for an age
of 1.8 Gyr using the reddenings and distances quoted above. 
We see that the slopes of the observed RGBs are somewhat steeper than those of the models. 
In addition, while the absolute location of the theoretical RGBs agree with that of NGC 1651,
it appears that they are too red as compared with the Hodge 4 RGB. Based on their locations 
relative to the isochrones, the two bright stars in NGC 1651 appear to be on the AGB.

\section{Conclusions}

We present deep JK CMDs for the LMC star clusters Hodge 4 and NGC 1651 revealing the
helium burning red clump stars for the first time in the JK passbands. From these diagrams, 
we draw the following conclusions. 

\noindent 1) The slope of the red giant branch indicates metallicities of $[Fe/H] = -0.17 \pm 0.04$
for Hodge 4 and $[Fe/H] = -0.07 \pm 0.10$ for NGC 1651. 

\noindent 2) By comparing existing optical CMDs that reveal the clusters' main sequence turnoffs
with theoretical isochrones, we find ages of $1.7 \pm 0.3$ Gyr for Hodge 4 and 
$1.8 \pm 0.3$ Gyr for NGC 1651.

\noindent 3) The Burstein \& Heiles (1982) and Schlegel et al. (1998) extinction maps suggest values
of $E(J-K) = 0.03 \pm 0.01$ and $E(J-K) = 0.06 \pm 0.01$ for Hodge 4 and NGC 1651.

\noindent 4) Using the GS01 K-band red clump luminosity calibration, we find distance moduli of
$(m-M)_0 = 18.52 \pm 0.17$ for Hodge 4 and $(m-M)_0 =18.55 \pm 0.12$ for NGC 1651 leading to
a mean distance for the LMC of $(m-M)_0 = 18.54 \pm 0.10$. The quoted errors are dominated
by uncertainties in the ages of these clusters. The contributions from errors in the cluster metallicities
and reddenings are significantly smaller.

\noindent 5) Comparison of the near-IR theoretical isochrones of Girardi et al. (2000) suggests that 
while the overall location of the RGBs matches the observed ones to within the errors, the slopes of 
the theoretical sequences are shallower than the observed RGBs.

\acknowledgments

The authors are grateful to Glenn Tiede for stimulating discussions and a careful reading of an
early version of this manuscript. The comments of an anonymous referee greatly improved the
quality of this paper. This research was supported by NSF CAREER grant AST-0094048.

\clearpage
\begin{deluxetable}{lcccccc}
\tablewidth{7in}
\tablecaption{Program Cluster Parameters} 
\tablehead{
\colhead{Cluster} & \colhead{$[Fe/H]$} & \colhead{Age (Gyr)} & 
\colhead{$K(RC)$} & \colhead{$M_K(RC)$} & \colhead{$E(J-K)$}  
& \colhead{$(m-M)_0$}}
\tablecolumns{6}
\startdata
Hodge 4     & $-0.17 \pm 0.04$ & $1.7 \pm 0.3$ & $16.90 \pm 0.02$ &  $-1.64 \pm 0.17$ & $0.03 \pm 0.01$ & $18.52 \pm 0.17$\\
NGC 1651 & $-0.07 \pm 0.10$ & $1.8 \pm 0.3$ & $17.03 \pm 0.02$ & $-1.56 \pm 0.12$ & $0.06 \pm 0.01$ & $18.55 \pm 0.12$\\
\enddata
\end{deluxetable}

\clearpage

\figcaption{Observed fields in the LMC clusters Hodge 4 (a) and NGC 1651 (b) in the $K_s$
filter passband. Each image is approximately 150" x 90" in size. North is up and east to the left.}

\figcaption{The points represent the standard star observations with the difference between
the exposure-time corrected instrumental magnitudes ($j$, $k_s$) and the standard values 
($J$, $K_s$) on the ordinate and the airmass on the abscissa. The solid lines are the fitted
relations given in Equations 1 and 2. The root-mean-square deviation of the points from the 
relations are 0.014 mag in $J$ and 0.013 mag in $K_s$.} 

\figcaption{Near-IR JK color-magnitude diagrams for Hodge 4 and NGC 1651. Both diagrams
reveal a well-populated red giant branch and core helium-burning red clump. In addition, NGC 1651
shows two bright stars that appear to be on the asymptotic giant branch.}

\figcaption{The left panels show our near-IR JK color-magnitude diagrams for Hodge 4 and 
NGC 1651 wherein the solid lines show the least squares fits to the clusters'
red giant branches. The filled symbols represent the fitted points. The rectangles, which are 0.2 mag
wide in $J-K$ and 0.8 mag in $K$, include the stars used to calculate the apparent K-band
magnitude of the red clump stars. The right panels display the luminosity function for
all stars in the CMDs.}

\figcaption{(a) The HST/WFPC2 photometry of Hodge 4 from Sarajedini (1998) is compared 
with the theoretical
isochrones of Girardi et al. (2000) for metallicities of Z=0.008 (upper panel) and Z=0.019 (lower panel).
(b) Same as (a) except that the HST/WFPC2 photometry of NGC 1651 from Mould et al. (1997) is
shown.}

\figcaption{Our near-IR photometry of Hodge 4 and NGC 1651 compared with the theoretical
giant branch sequences from Girardi et al. (2000) for metallicities of Z=0.008 (left line) and
Z=0.019 (right line).}


\end{document}